\documentclass[a4paper,12pt]{article}
\usepackage[english]{babel}
\usepackage{graphicx,color}
\usepackage{ifpdf}
\usepackage{amsmath,amssymb,amsfonts,bm,euscript,amscd,ae,amsthm,exscale,amsopn,amstext}
\usepackage{ifthen}
\usepackage[T1]{fontenc}
\usepackage[cp1251]{inputenc}
\usepackage[active]{srcltx}

\newcommand{\tr}{\mathrm{Tr}}

\newcommand{\vac}{\left|\mathrm{VAC}\right\rangle}
\newcommand{\ivac}{\left\langle\mathrm{VAC}\right|}

\title{Thermodynamical Quantum Gravity}

\author{\textbf{Lukasz Andrzej Glinka}\footnote{E-mail address: laglinka@gmail.com, lukaszglinka@wp.eu}}

\date{\empty}

\begin{document}
\maketitle

\begin{abstract}
The canonically quantized $3+1$ General Relativity with the global one dimensionality conjecture defines the model, which dimensionally reduced and secondary quantized yields the one-dimensional quantum field theory wherein the generic one-point correlations create a boson mass responsible for quantum gravity. In this paper, this simple model is developed in a wider sense. We propose to consider the thermodynamics of space quanta, constructed \emph{ab initio} from the entropic formalism, as the quantum gravity phenomenology.\\

\noindent \textbf{Keywords:} quantum gravity phenomenology, quantum geometrodynamics, thermodynamics, space quanta, entropy of graviton
\end{abstract}
\newpage
\section{Introduction}

Both the theory and phenomenology of quantum gravity possess the most fundamental meaning for the contemporary theoretical physics, wherein the quantum nature of gravitation still remains an experimentally non verified field. Perhaps, the theory of quantized gravitational fields will able to predict unknown information and open the way for the new cosmology. The efforts of many generations of researchers who have worked on the constructive and consistent formulation of quantum gravity unquestionably have given a significant contribution to science, but still are insufficient to give an insight into the quantum nature of gravitational phenomena. In this a great success, however, understanding the physical role of quantum gravity seems to be still very distant and intriguing perspective, Cf. the citations in the Ref. \cite{qg}.

In this paper, we discuss the subsequent implication following form the simple model of quantum gravity recently considered by the author \cite{glinka}, which has the strict roots in the primarily formulated generic approach to quantum cosmology \cite{gli}. Recall that the model was constructed within the Wheeler--DeWitt theory, well-known as the quantum geometrodynamics, throughout taking into account the global one-dimensional conjecture. The conjecture states that a geometrodynamical wave function is dependent on the one dimension only, that is a determinant of an embedded space metric. It follows from the assumption that Matter fields are functionals of this dimension only and, consequently, reduces quantum geometrodynamics to the superspace strata. Through application of the reduction, the resulting quantum gravity is formulated through the Dirac equation with the Euclidean Clifford algebra $\mathcal{C}\ell_{1,1}(\mathbb{R})$, and making use of the appropriate diagonalization, the equation is quantized in the Fock space. The received one-dimensional quantum field theory defines the quantum gravity, wherein quantum correlations yield the mass of a boson responsible for quantum nature of gravitation.

Nevertheless, the investigated simple model of quantum gravity can be still developed. This paper gives one of its possible physical implications, that is the thermodynamics of space quanta, where through the space quanta we understand the quantum states of a three-dimensional space embedded in the four dimensional Lorentzian space-time. The Fock space formulation offered by quantum field theory, gives a possibility to consider the density matrix related to the model, and build the formal \textit{ab initio} thermodynamics on the basis of the entropic formalism. As the leading example, we are discussing the one-particle approximation. Entropy and energy are calculated, and their appropriate renormalization is simply performed. In result, we obtain the second order Eulerian system, and all thermodynamical quantities are calculated in the framework of the standard statistical mechanics through application of the first principles only. In consequence, we receive the model which possesses a natural interpretation of quantum gravity phenomenology.

Structurally the paper is organized as follows. The preliminary section \ref{sec:1} briefly presents the simple model of quantum gravity. Section \ref{sec:4} is devoted to the development of the model, that is the thermodynamics of space quanta. In the final section \ref{sec:5}, the new results are briefly discussed.

\section{The simple quantum gravity model}\label{sec:1}
Let us summarize the quantum gravity model \cite{glinka}. Regarding General Relativity \cite{gr}, a space-time is a four-dimensional Lorentizan (pseudo-Riemannian) manifold $(M,g)$ with a metric $g_{\mu\nu}$ satisfying the Einstein field equations
\begin{equation}\label{feq}
R_{\mu\nu}-\dfrac{1}{2}g_{\mu\nu}{^{(4)}}\!R+g_{\mu\nu}\Lambda=3T_{\mu\nu},
\end{equation}
where the units $c=8\pi G/3=1$ are used, $R_{\mu\nu}$ is the second fundamental form, ${^{(4)}}\!R$ is the scalar curvature, $\Lambda$ is cosmological constant, and $T_{\mu\nu}$ is a stress-energy tensor arising from a Matter fields Lagrangian $\mathcal{L}_\psi$
\begin{equation}
   T_{\mu\nu}=-\dfrac{2}{\sqrt{{-g}}}\dfrac{\delta S_\phi}{\delta g^{\mu\nu}}\quad,\quad S_\phi=\int d^4x \sqrt{{-g}}\mathcal{L}_\psi.
 \end{equation}
For $M$ closed, having a boundary $(\partial M,h)$ with an induced metric $h_{ij}$ and the Gauss curvature tensor $K_{ij}$, (\ref{feq}) are the equations of motion for the Einstein--Hilbert action supplemented by the York--Gibbons--Hawking term \cite{ygh}
\begin{equation}\label{eh0}
S[g]=\int_{M}d^4x\sqrt{-g}\left\{-\dfrac{1}{6}{^{(4)}}R+\dfrac{\Lambda}{3}\right\}+S_\phi-\dfrac{1}{3}\int_{\partial M}d^3x\sqrt{h}K,
\end{equation}
where $K=h^{ij}K_{ij}$. One can parameterize a metric by the $3+1$ splitting \cite{adm}
\begin{equation}\label{dec}
g_{\mu\nu}=\left[\begin{array}{cc}-N^2+N_iN^i&N_j\\N_i&h_{ij}\end{array}\right]\quad,\quad N^i=h^{ij}N_j\quad,\quad h_{ik}h^{kj}=\delta_i^j,
\end{equation}
which for stationary $\phi$ arises by a timelike Killing vector field and global spacelike foliation $t=\mathit{const}$ on $M$, and satisfies the Nash embedding theorem \cite{nash}. With this (\ref{eh0}) takes the Hamilton form $S=\int dt L$ with the Lagrangian
\begin{eqnarray}\label{gd}
\!\!\!\!\!\!\!\!\!\!&&L=\int_{\partial M} d^3x\left\{\pi\dot{N}+\pi^i\dot{N_i}+\pi^{ij}\dot{h}_{ij}+\pi_\phi\dot{\phi}-NH-N_iH^i\right\},\\
\!\!\!\!\!\!\!\!\!\!&&\pi_\phi=\frac{\partial L_\phi}{\partial \dot{\phi}}\quad,\quad\pi=\frac{\partial L}{\partial \dot{N}}\quad,\quad\pi^i=\frac{\partial L}{\partial \dot{N_i}},\\
\!\!\!\!\!\!\!\!\!\!&&\pi^{ij}=\frac{\partial L}{\partial\dot{h}_{ij}}=\sqrt{h}\left(h^{ij}K-K^{ij}\right)\quad,\quad\dot{h}_{ij}=N_{i|j}+N_{j|i}-2NK_{ij},\label{con0}\\
\!\!\!\!\!\!\!\!\!\!&&H=\sqrt{h}\left\{K^2-K_{ij}K^{ij}+{^{(3)}}R-2\Lambda-6\varrho\right\}\quad,\quad H^i=-2\pi^{ij}_{~;j},
\end{eqnarray}
where $\varrho=n^\mu n^\nu T_{\mu\nu}$, $n^\mu=(1/N)\left[1,-N^i\right]$, ${^{(3)}}R=h^{ij}R_{ij}$. Time-preservation of the primary constraints \cite{dir,dew} yields the secondary ones
\begin{equation}
  \pi\approx0\quad,\quad\pi^i\approx0 \longrightarrow H\approx0\quad,\quad H^i\approx0\label{const}
\end{equation}
called the Hamiltonian (scalar) and the diffeomorphism (vector) constraint. Vector constraint reflects spatial diffeoinvariance, scalar one is dynamical. Regarding DeWitt \cite{dew} $H^i$ generates the diffeomorphisms $\widetilde{x}^i=x^i+\xi^i$
\begin{eqnarray}
i\left[h_{ij},\int_{\partial M}H_{a}\xi^a d^3x\right]&=&-h_{ij,k}\xi^k-h_{kj}\xi^{k}_{~,i}-h_{ik}\xi^{k}_{~,j}~~,\\
i\left[\pi^{ij},\int_{\partial M}H_{a}\xi^a d^3x\right]&=&-\left(\pi^{ij}\xi^k\right)_{,k}+\pi^{kj}\xi^{i}_{~,k}+\pi^{ik}\xi^{j}_{~,k}~~,
\end{eqnarray}
and the first-class constraints algebra holds
\begin{eqnarray}
  &&i\left[H_i(x),H_j(y)\right]=\int_{\partial M}H_{a}c^a_{ij}d^3z\quad,\quad i\left[H(x),H_i(y)\right]=H\delta^{(3)}_{,i}(x,y),~~~~~~\label{com2}\\
  &&i\left[\int_{\partial M}H\xi_1d^3x,\int_{\partial M}H\xi_2d^3x\right]=\int_{\partial M}H^a\left(\xi_{1,a}\xi_2-\xi_1\xi_{2,a}\right)d^3x.~~~~~~\label{com3}
\end{eqnarray}
Here $H_i=h_{ij}H^j$, and $c^a_{ij}=\delta^a_i\delta^b_j\delta^{(3)}_{,b}(x,z)\delta^{(3)}(y,z)-(i\leftrightarrow j,x\leftrightarrow y)$ are the structure constants of the spatial diffeomorphism group. The canonical quantization \cite{dir,fad}
\begin{eqnarray}\label{dpq}
&i\left[\pi^{ij}(x),h_{kl}(y)\right]=\dfrac{1}{2}\left(\delta_{k}^{i}\delta_{l}^{j}+\delta_{l}^{i}\delta_{k}^{j}\right)\delta^{(3)}(x,y),&\\
&i\left[\pi^i(x),N_j(y)\right]=\delta^i_j\delta^{(3)}(x,y)\quad,\quad i\left[\pi(x),N(y)\right]=\delta^{(3)}(x,y),&
\end{eqnarray}
applied to the Hamiltonian constraint into the Hamilton--Jacobi form \cite{hj}
\begin{equation}\label{con}
G_{ijkl}\pi^{ij}\pi^{kl}-\sqrt{h}\left({^{(3)}}R-2\Lambda-6\varrho\right)=0,
\end{equation}
where $G_{ijkl}$ is the DeWitt metric on the Wheeler superspace \cite{sup}
\begin{equation}
G_{ijkl}\equiv\dfrac{1}{2\sqrt{h}}\left(h_{ik}h_{jl}+h_{il}h_{jk}-h_{ij}h_{kl}\right),
\end{equation}
yields the Wheeler--DeWitt equation \cite{dew,whe,qgr1}
\begin{equation}\label{wdw}
\left\{G_{ijkl}\dfrac{\delta^2}{\delta h_{ij}\delta h_{kl}}+h^{1/2}\left({^{(3)}}R-2\Lambda-6\varrho\right)\right\}\Psi[h_{ij},\phi]=0.
\end{equation}
Other first-class reflect diffeoinvariance of a wave function $\Psi[h_{ij},\phi]$
\begin{equation}
  \pi\Psi[h_{ij},\phi]=0,~~\pi^i\Psi[h_{ij},\phi]=0,~~H^i\Psi[h_{ij},\phi]=0.
\end{equation}

The author's simple model reduces the superspace through the assumption that Matter fields are one-dimensional functionals
\begin{equation}
\phi=\phi[h]\quad,\quad h=\dfrac{1}{3}\varepsilon^{ijk}\varepsilon^{abc}h_{ia}h_{jb}h_{kc}\quad,
\end{equation}
where $\varepsilon$ is the Levi-Civita tensor. Consequently, the conjectured wave functions
\begin{equation}\label{GOD}
  \Psi[h_{ij},\phi]\rightarrow\Psi[h],
\end{equation}
satisfy the global one-dimensional evolution
\begin{equation}\label{wdw1}
\left\{G_{ijkl}\dfrac{\delta^2}{\delta h_{ij}\delta h_{kl}}+h^{1/2}\left({^{(3)}}R-2\Lambda-6\varrho[h]\right)\right\}\Psi[h]=0.
\end{equation}
Assumption (\ref{GOD}) is analogous to the generic model \cite{gli}, but the theory (\ref{wdw1}) particularly holds for nonhomogeneous isotropic quantum cosmologies.

Considering the Jacobi rule for differentiation of a determinant \cite{gr} together with the $3+1$ splitting (\ref{dec}) one obtains
\begin{equation}\label{hij}
  \delta g = gg^{\mu\nu}\delta g_{\mu\nu} \longrightarrow N^2\delta h=N^2hh^{ij}\delta h_{ij},
\end{equation}
which reduces the differential operator in (\ref{wdw1})
\begin{equation}\label{reds}
  \dfrac{\delta}{\delta h_{ij}}\Psi[h]= hh^{ij}\dfrac{\delta}{\delta h}\Psi[h] \longrightarrow G_{ijkl}\dfrac{\delta^2}{\delta h_{ij}\delta h_{kl}}\Psi[h]=-\dfrac{3}{2}h^{3/2}\dfrac{\delta^2}{\delta h^2}\Psi[h],
\end{equation}
and yields the simple one-dimensional quantum gravity model
\begin{equation}\label{kgf}
\left(\dfrac{\delta^2}{\delta{h^2}}-m^2\right)\Psi=0\quad,\quad m^2=\dfrac{2}{3h}\left({^{(3)}}R-2\Lambda-6\varrho[h]\right).
\end{equation}
where $m$ is the mass of the classical field $\Psi[h]$. In fact, (\ref{kgf}) is a field-theoretic equation of motion $\delta S[\Psi]/\delta\Psi=0$ for the Euclidean action
\begin{equation}
  S[\Psi]=\int \delta h L[\Psi,\Pi_\Psi]\quad,\quad L=\dfrac{1}{2}\Pi_\Psi^2+\dfrac{m^2}{2}\Psi^2,
\end{equation}
where $\Pi_\Psi=\dfrac{\delta \Psi}{\delta h}$ is conjugate momentum which allows rewrite (\ref{kgf}) in two-component model in the Dirac equation form
\begin{equation}\label{dira}
 \left(i\gamma\dfrac{\delta}{\delta h}-M\right)\Phi=0\quad,\quad\Phi=\left[\begin{array}{c}\Pi_\Psi\\ \Psi\end{array}\right]\quad,\quad M=\left[\begin{array}{cc}
1&0\\0&-m^{2}\end{array}\right],
\end{equation}
with the Euclidean Clifford algebra $\mathcal{C}\ell_{1,1}(\mathbb{R})$ \cite{euc}
\begin{equation}\label{cliff}
  \gamma=\left[\begin{array}{cc}0&-i\\i&0\end{array}\right]\quad,\quad\gamma^2=I\quad,\quad \left\{\gamma,\gamma\right\}=2\delta_E\quad,\quad\delta_E=\left[\begin{array}{cc}1&0\\0&1\end{array}\right],
\end{equation}
having a two-dimensional complex representation. Restricting to $Pin_{1,1}(\mathbb{R})$ yield a two-dimensional spin representations; to $Spin_{1,1}(\mathbb{R})$ splits it onto a sum of two one-dimensional Weyl representations. $\mathcal{C}\ell_{1,1}(\mathbb{R})$ decomposes into a direct sum of two isomorphic central simple algebras or a tensor product
\begin{eqnarray}
\!\!\!\!\!\!\!\!\!\!&&\mathcal{C}\ell_{1,1}(\mathbb{R})=\mathcal{C}\ell^+_{1,1}(\mathbb{R})\oplus \mathcal{C}\ell^-_{1,1}(\mathbb{R})=\mathcal{C}\ell_{2,0}(\mathbb{R})\otimes\mathcal{C}\ell_{0,0}(\mathbb{R}),~~~~~~\\
\!\!\!\!\!\!\!\!\!\!&&\mathcal{C}\ell_{1,1}(\mathbb{R})\cong\mathbb{R}(2)\quad,\quad\mathcal{C}\ell^{\pm}_{1,1}(\mathbb{R})=\dfrac{1\pm\gamma}{2}\mathcal{C}\ell_{1,1}(\mathbb{R})\cong\mathbb{R}\quad,\quad\mathcal{C}\ell_{0,0}(\mathbb{R})\cong\mathbb{R}.~~~~~~
\end{eqnarray}

The Dirac equation (\ref{dira}) can be rewritten in the dynamical Fock reper $\mathfrak{B}$
\begin{eqnarray}\label{sqx}
  &&\mathbf{\Phi}=\mathbb{Q}\mathfrak{B},~~\mathbb{Q}=\left[\begin{array}{cc}1/\sqrt{2|m|}&1/\sqrt{2|m|}\\
-i\sqrt{|m|/2}&i\sqrt{|m|/2}\end{array}\right],\\
&&\mathfrak{B}=\left\{\left[\begin{array}{c}\mathsf{G}[h]\\
\mathsf{G}^{\dagger}[h]\end{array}\right]:\left[\mathsf{G}[h'],\mathsf{G}^{\dagger}[h]\right]=\delta\left(h'-h\right), \left[\mathsf{G}[h'],\mathsf{G}[h]\right]=0\right\}.\label{db}
\end{eqnarray}
Determining a reper $\mathfrak{F}$ by the diagonalization due to the Bogoliubov transformation and the Heisenberg equations of motion
\begin{eqnarray}
\mathfrak{F}=\left[\begin{array}{cc}u&v\\v^{\ast}&u^{\ast}\end{array}\right]\mathfrak{B}\quad,\quad\dfrac{\delta\mathfrak{F}}{\delta h}=\left[\begin{array}{cc}-i\Omega&0\\0&i\Omega\end{array}\right]\mathfrak{F},\label{pr3}
\end{eqnarray}
where $|u|^2-|v|^2=1$, $u$, $v$, $\Omega$ are functionals of $h$, one obtains
\begin{equation}\label{bcof}
  \dfrac{\delta\mathbf{b}}{\delta h}=\mathbb{X}\mathbf{b}\quad,\quad\mathbf{b}=\left[\begin{array}{c}u\\v\end{array}\right]\quad,\quad \Omega\equiv0.
\end{equation}
Therefore, $\mathfrak{F}$ is the Fock initial data static reper ($I$) with stable vacuum
\begin{equation}\label{in}
\mathfrak{F}=\left\{\left[\begin{array}{c}\mathsf{G}_I\\
\mathsf{G}^{\dagger}_I\end{array}\right]: \left[\mathsf{G}_I,\mathsf{G}^{\dagger}_I\right]=1, \left[\mathsf{G}_I,\mathsf{G}_I\right]=0\right\}\quad,\quad \mathsf{G}_I\vac=0,
 \end{equation}
and integrability of (\ref{bcof}) can be done in the superfluid parametrization
\begin{eqnarray}
u=\dfrac{\mu+1}{2\sqrt{\mu}}e^{i\theta}\quad,\quad v=\dfrac{\mu-1}{2\sqrt{\mu}}e^{-i\theta}\quad,\quad \theta=m_I\int_{h_I}^{h}\mu'\delta h',\label{sup2}
\end{eqnarray}
where $\mu\equiv\mu[h]$, $\mu'=\mu[h']$ is a mass scale. In result, one obtains the solution
\begin{equation}\label{phi}
\mathbf{\Phi}=\mathbb{Q}\mathbb{G}\mathfrak{F}\quad,\quad \mathbb{G}=\left[\begin{array}{cc}u^\star&-v^\star\\-v&u\end{array}\right],\end{equation}
and establishes the field operator and the generic one-point correlator
\begin{eqnarray}\label{field}
  \mathbf{\Psi}=\frac{1}{\sqrt{2m_I}}\left(\dfrac{e^{-i\theta}}{2\mu}\mathsf{G}_I+\dfrac{e^{i\theta}}{2\mu}\mathsf{G}_I^\dagger\right)\quad,\quad\ivac\mathbf{\Psi}^\dagger[h]\mathbf{\Psi}[h]\vac=\dfrac{1}{\mu^2},
\end{eqnarray}
where the quantum correlator was normalized to unity in $h_I$.

\section{The thermodynamics}\label{sec:4}
The static Fock reper formulation defines the self-consistent concept of space quanta, which are the quantized fields associated with an three-dimensional space embedded in the four-dimensional space-time. Thermodynamical equilibrium corresponding to the static quantum field theory, allows to make use of first principles of statistical mechanics \cite{ker}, and formulate the \emph{ab initio} thermodynamics of space quanta. Let us test the one-particle density matrix approximation.
\subsection{One-particle density matrix. Entropy and energy}
In the one-particle approximation, the density operator $\mathsf{D}$ is equivalent to an occupation number operator, whereas the thermodynamical equilibrium is determined with respect to the static reper. Therefore, the equilibrium one-particle density matrix $\mathbb{D}$ is given by the Von Neumann--Heisenberg picture
\begin{eqnarray}
\mathsf{D}&=&{\mathsf{G}}^{\dagger}{\mathsf{G}}=\mathfrak{F}^{\dagger}\mathbb{D}\mathfrak{F},\\
\mathbb{D}&=&\dfrac{1}{4\mu}\left[\begin{array}{cc}(\mu+1)^2&1-\mu^2\\
  1-\mu^2&(\mu-1)^2\end{array}\right]\label{den}.
\end{eqnarray}
Apparently $\det\mathbb{D}=0$, that is the thermodynamics in the one-particle approximation is not invertible. Employing (\ref{den}), one obtains the occupation number
\begin{equation}
  N=\dfrac{\tr\left(\mathbb{D}^2\right)}{\tr\mathbb{D}}=\dfrac{\mu^2+1}{2\mu},
\end{equation}
and the entropy can be derived from its basic definition
\begin{eqnarray}
  S=-\dfrac{\tr(\mathbb{D}\ln\mathbb{D})}{\tr\mathbb{D}}=\sum_{n=1}^{\infty}\sum_{k=1}^n\dfrac{(-1)^{k}}{n}\binom{n}{k-1}S_k,\label{entropia0}
\end{eqnarray}
where $\binom{n}{m}$ are the Newton binomial symbols, and
\begin{equation}
  S_k=\dfrac{\tr(\mathbb{D}^{k})}{\tr\mathbb{D}}=N^{k-1},
\end{equation}
are cluster entropies. The series (\ref{entropia0}) converges for the spectral radius values
\begin{equation}
\rho(\mathbb{D}-\mathbb{I})<1\Longrightarrow\mu\in(1;2+\sqrt{3}),
\end{equation}
or equivalently for $m\in(1;2+\sqrt{3})m_I$, with the result
\begin{equation}
  S=-\dfrac{\zeta(1)}{2}\left(\dfrac{\mu^2-1}{\mu^2+1}\right)^2-\dfrac{\mu^4+6\mu^2+1}{(\mu^2+1)^2}\ln\dfrac{(\mu-1)^2}{2\mu},\label{entropia}
\end{equation}
where $\zeta(s)=\sum_{n=1}^{\infty}\dfrac{1}{n^s}$ is the Riemann zeta function; $\zeta(1)$ is formally infinite.

Interestingly, a straightforward application of the Hagedorn hadronization formula $m\sim T_H$ \cite{hag}, where $m$ is the mass of the system, establishes the hadronized temperature
\begin{equation}
  \dfrac{T_H}{T_I}=\mu,
\end{equation}
where $k_BT_I=m_Ic^2$. By the relation $\langle m \rangle c^2 \sim k_B\langle T_H\rangle$ one obtains the averaged hadronized temperature normalized to $T_I$ value
\begin{equation}\label{srednia}
\left\langle \dfrac{T_H}{T_I}\right\rangle=\langle\mu\rangle=\dfrac{1+\sqrt{3}}{2}\approx 2.732,
\end{equation}
and, consequently, one can establish the ratio
\begin{equation}
\dfrac{\left\langle T_H\right\rangle}{T_H}\in\left(\dfrac{\sqrt{3}-1}{2},\dfrac{\sqrt{3}+1}{2}\right)\approx\left(0.366,1.366\right).
\end{equation}
Defining anisotropy as $\Delta T_H=\left\langle T_H\right\rangle-T_H$ one derives
\begin{equation}
  \dfrac{\Delta T_H}{T_H}\in\left(\dfrac{\sqrt{3}-3}{2},\dfrac{\sqrt{3}-1}{2}\right)\approx(-0.633;0.366),
\end{equation}
and, therefore, the averaged anisotropy is
\begin{equation}
  \left\langle\dfrac{\Delta T_H}{T_H}\right\rangle=0.5.
\end{equation}
By the first approximation, the relation (\ref{srednia}) can be identified with the background temperature, \emph{e.g.} for $T_I\sim1K$ this is exactly the averaged Cosmic Background Radiation temperature. Next approximations of the density matrix, or fuzzing of the interval $\mu\in(1;2+\sqrt{3})$, will give the next orders.

In the static reper, the Hamiltonian matrix $\mathbb{H}$ of the system equals
\begin{eqnarray}
  \mathsf{H}&=&\dfrac{m}{2}\left(\mathsf{G}^\dagger\mathsf{G}+\mathsf{G}\mathsf{G}^\dagger\right)=\mathfrak{F}^\dagger\mathbb{H}\mathfrak{F},\\
  \mathbb{H}&=&\dfrac{m_I}{4}\left[\begin{array}{cc} 1+\mu^2& 1-\mu^2\\ 1-\mu^2& 1+\mu^2\end{array}\right],\label{shm}
\end{eqnarray}
and has a discrete spectrum for the fixed value of a mass scale
\begin{equation}
\mathrm{Spec}~\mathbb{H}=\left\{\dfrac{m_I}{2}\mu^2,\dfrac{m_I}{2}\right\}.
\end{equation}
The internal energy calculated from the Hamiltonian matrix (\ref{shm}) is
\begin{equation}\label{int}
  U=\dfrac{\tr(\mathbb{D}\mathbb{H})}{\tr\mathbb{D}}=\dfrac{m_I}{4}(\mu^2+1).
\end{equation}
The Hamiltonian matrix $\mathbb{H}$, however, consists of a constant term $\mathbb{H}_{I}$
\begin{equation}
  \mathbb{H}_I=\dfrac{m_I}{4}\left[\begin{array}{cc} 1& 1\\ 1& 1\end{array}\right]
\end{equation}
which can be eliminated through a simple renormalization
\begin{equation}
  \mathbb{H}\rightarrow\mathbb{H}'=\mathbb{H}-\mathbb{H}_I=\dfrac{m_I}{4}\left[\begin{array}{cc} \mu^2&-\mu^2\\-\mu^2&\mu^2\end{array}\right].
\end{equation}
Consequently, the renormalized Hemiltonian spectrum is
\begin{equation}
  \mathrm{Spec}~\mathbb{H}'=\left\{\dfrac{m_I}{2}\mu^2,0\right\},
\end{equation}
and straightforward computation of the renormalized internal energy yields the following result
\begin{equation}\label{int1}
  U'=\dfrac{\tr(\mathbb{D}\mathbb{H}')}{\tr\mathbb{D}}=\dfrac{m_I}{4}\mu^2\equiv U-U_I,
\end{equation}
where $U_I=\dfrac{m_I}{4}$, which has the Eulerian homogeneity of degree two
\begin{equation}
U'[\alpha\mu]=\alpha^2U'[\mu].
\end{equation}
In this manner, thermodynamics of space quanta can be formulated in the way typical for the Eulerian systems.

\subsection{\emph{Ab initio} thermodynamics of space quanta}
Three elementary thermodynamical quantities, that are the occupation number $N$, internal energy $U$, and entropy $S$, just have been derived, and now one can conclude the remaining formalities. Actually, the entropy (\ref{entropia}) is infinite through the presence of a formal infinity $\zeta(1)$. Straightforward calculation shows that temperature $T=\delta U/\delta S$ arising from the entropy (\ref{entropia}) is dependent on $\zeta(1)$ and the initial data mass $m_I$. Obtained quantity has the finite limit if and only if the initial data mass is scaled $m_I \rightarrow m_I\zeta(1)$. Since the boson mass $m$ is related to length $l$ by $m\sim1/l$, the limit $m_I\rightarrow\infty$ corresponds with a point object $l_I\rightarrow0$.

However, scaling of the initial data is not a good physical procedure, that is has not a well-defined physical meaning. It can be shown that the entropy renormalization $S\rightarrow -S/\zeta(1)$ in the formal limit $\zeta(1)\rightarrow\infty$ gives the equivalent result for the thermodynamics with no using the initial data scaling. The renormalization corresponds to an initial quantum state of an embedded space being a point, and yields a perfect accordance with the second law of thermodynamics
\begin{equation}
  S\longrightarrow S'=\lim_{\zeta(1)\rightarrow\infty}\dfrac{-S}{\zeta(1)}=\dfrac{1}{2}\left(\dfrac{\mu^2-1}{\mu^2+1}\right)^2\geqslant0.\label{sren}
\end{equation}
Calculating the temperature $T$ of space quanta, one obtains
\begin{equation}
  T=\dfrac{\delta U'}{\delta S'}=m_I\dfrac{(\mu^2+1)^3}{8(\mu^2-1)},
\end{equation}
and one sees that initially, that is for $\mu=1$, this temperature is infinite, what corresponds with the Hot Big Bang phenomenon. After reaching this points, the system is cooled right up until the mass scale value $\mu_{PT}=\sqrt{2}\approx1.414$, and then is warmed. In fact, $\mu_{PT}$ is the phase transition point, namely, the energetic heat capacity $C_{U}$, having the form
\begin{equation}\label{heat}
  C_U=T\dfrac{\delta S'}{\delta T}=\dfrac{\delta U'}{\delta T}=\dfrac{(\mu^2-1)^2}{(\mu^2-2)(\mu^2+1)^2},
\end{equation}
possesses the singularity in the point $\mu_{PT}$. The generalized law of equipartition $\delta U/\delta T=f/2$ establishes the degrees of freedom $f$ number
\begin{equation}
  f=2C_{U}.
\end{equation}
The Helmholtz free energy \mbox{$F=U'-TS'$}, that is
\begin{equation}
  F=-\dfrac{m_I}{16}(\mu^4-4\mu^2-1),
\end{equation}
is finite for finite values of $m_I$, increases since $\mu=1$ till $\mu_{PT}$, and then decreases. Therefore, the thermal equilibrium point is $\mu_{eq}=1$, that is the Hot Big Bang. In the region where the mass scale is $1\leqslant\mu<\mu_{PT}$, the mechanical isolation is absent, but this is after the phase transition $\mu>\mu_{PT}$. Calculating the chemical potential
\begin{equation}
  \omega=\dfrac{\delta F}{\delta N}=-m_I\dfrac{\mu^3(\mu^2-2)}{2(\mu^2-1)},\label{chem}
\end{equation}
one sees that, in the point of equilibrium $\mu_{eq}$, its value diverges, whereas in the point of phase transition $\mu_{PT}$ its value vanishes. Making use of the formula (\ref{chem}), together with the occupation number $N$ and the Helmholtz free energy $F$, yields the appropriate free energy defined by the Landau grand potential $\Omega$
\begin{equation}
  \Omega=F-\omega N=m_I\dfrac{3\mu^6+\mu^4-11\mu^2-1}{16(\mu^2-1)},
\end{equation}
and, consequently, the corresponding Massieu--Planck free entropy $\Xi$ can be also derived
\begin{equation}
  \Xi=-\dfrac{\Omega}{T}=-\dfrac{3\mu^6+\mu^4-11\mu^2-1}{2(\mu^2+1)^3},
\end{equation}
whereas the grand partition function $Z$ is established as
\begin{eqnarray}\label{gpf}
  Z=e^{\Xi}=\exp\left\{-\dfrac{3\mu^6+\mu^4-11\mu^2-1}{2(\mu^2+1)^3}\right\}.
\end{eqnarray}
The second order Eulerian homogeneity yields the equation of state $PV/T =\ln Z$ and determines the product of pressure $P$ and volume $V$ as
\begin{equation}\label{eos}
  PV=-\Omega,
\end{equation}
and, therefore, together with the Gibbs--Duhem equation $V\delta P=S'\delta T+N\delta\omega$ allows to establish the pressure
\begin{equation}
  |P|=\exp\left\{-\int\left(S+N\dfrac{\delta \omega}{\delta T}\right)\dfrac{\delta T}{\Omega}\right\}.
\end{equation}
Similarly, the first law of thermodynamics, $-\delta\Omega=S'\delta T+P\delta V+N\delta \omega$, and the equation of state (\ref{eos}) determine the volume $|V|=|\Omega|/|P|$, which through positiveness is $V=|V|$. Regarding (\ref{eos}), the pressure $P=|P|$ for $\Omega=-|\Omega|<0$, and $P=-|P|$ for $\Omega=|\Omega|>0$ and, therefore, one has
\begin{eqnarray}
  P=\left\{\begin{array}{ll}\dfrac{m_I^7a_0}{\mu^2-1}\dfrac{(\mu^2+a_2)^{b_2+1}}{(\mu^2+a_3)^{b_3-1}}|\mu^2-a_1|^{b_1+1}&,~\mathrm{iff}~1\leqslant\mu\leqslant\sqrt{a_1}\vspace*{10pt}\\
  \dfrac{-m_I^7a_0}{\mu^2-1}\dfrac{(\mu^2+a_2)^{b_2+1}}{(\mu^2+a_3)^{b_3-1}}|\mu^2-a_1|^{b_1+1}&,~\mathrm{iff}~\sqrt{a_1}\leqslant\mu\leqslant2+\sqrt{3}\end{array}\right.
\end{eqnarray}
where $a_0\approx6.676\cdot10^6$ and
\begin{eqnarray}
&a_1\approx1.802~,~a_2\approx0.090~,~a_3\approx2.046~,&\\
&b_1\approx0.014~,~b_2\approx0.410~,~b_3\approx1.092~.&
\end{eqnarray}
For the mass scale is the region $1\leqslant\mu<\sqrt{a_1}$, the pressure $P$ decreases from a positive infinity to zero, vanishes in the point $\mu=\sqrt{a_1}\approx1.343$, and decreases from zero to a negative infinity whenever $\sqrt{a_1}<\mu\leqslant2+\sqrt{3}$. Regarding the relation $V=|\Omega|/|P|$, the volume $V$ is a fixed parameter and can be established as follows
\begin{equation}\label{vol}
V=\dfrac{1}{16a_0m_I^6}\dfrac{1}{|\mu^2-a_1|^{b_1}}\dfrac{(\mu^2+a_3)^{b_3}}{(\mu^2+a_2)^{b_2}}.
\end{equation}

Equivalently, the thermodynamics of space quanta can be expressed through the size scale $\lambda=\dfrac{1}{\mu}$. There are the relations joining both the scales with an occupation number
\begin{eqnarray}
  \lambda=N\left(1\mp\sqrt{\strut{1-\dfrac{1}{N^2}}}\right)\quad,\quad \mu=N\left(1\pm\sqrt{\strut{1-\dfrac{1}{N^2}}}\right),
\end{eqnarray}
that in the limit of infinite $N$ are equal
\begin{eqnarray}
  \lambda_{N=\infty}=\left\{0,\infty\right\}\quad,\quad \mu_{N=\infty}=\left\{\infty,0\right\},
\end{eqnarray}
and, therefore, there are two possible asymptotic behaviors. The first case, that is $\lambda=0$ and $\mu=\infty$, can be interpreted with the Hot Big Bang from a black hole, whereas the second one, that is $\lambda=\infty$ and $\mu=0$, is an infinite massless classical space-time.

\section{Discussion}\label{sec:5}

In this paper, we have presented the subsequent implication of the simple model of quantum gravity recently proposed by the author \cite{glinka}. This algorithm has yielded a constructive, consistent and, moreover, plausible quantum gravity phenomenology, given through thermodynamics which in the discussed state of affairs has been described the statistical behaviour of space quanta. Certainly, the model applies to all $3+1$ general relativistic space-times which satisfy the Mach principle, or, in other words, are isotropic. Their crucial importance for elementary particle physics, cosmology and high energy astrophysics is experimentally confirmed; one can say that these are the phenomenological space-times.

As the example of the \emph{ab initio} formulated thermodynamics, we have employed the one-particle approximation of density matrix. The renormalization method was applied for entropy and the Hamiltonian matrix, and has yielded the second order Eulerian homogeneity property. The Landau grand potential $\Omega$ and the Massieu--Planck free entropy $\Xi$ were used in the phenomenological description. Grand partition function $Z$ was established. Thermodynamical volume $V$ was determined as fixed parameter. Other thermodynamical potentials were derived in the framework of the entropic formalism, that is in accordance with both the first and  second principles of thermodynamics. The physical information following from the thermodynamics of space quanta is the crucial point of the presented construction. Actually, the approach of this paper differs from the other ones, Cf. the Ref. \cite{th}, through a straightforward derivation of the \emph{ab initio} quantum gravity phenomenology.

In our opinion, studying of the specific physical phenomena through the proposed approach seems to be the most important prospective arising from the thermodynamics of space quanta. From the experimental point of view, the presented considerations possess a clear potential usefulness, because of the bosonic systems are common in high energy physics. A reader interested in more insightful developments is advised refer to the author's monograph \cite{glinkamon}.

\end{document}